\documentclass[11pt,twoside]{article}
\usepackage{cal05}
\usepackage{graphics}
\usepackage{epsf}

\contact{Roelof de Jong}
\email{dejong@stsci.edu}

\begin{document}

%
%
%

\title{NICMOS Status}

%
%
%

\author{Roelof S.\ de Jong, Santiago Arribas, Elizabeth Barker, Louis
  E.\ Bergeron, Ralph C.\ Bohlin, Daniela Calzetti, Ilana Dashevsky,
  Mark Dickinson, Anton M.\ Koekemoer, Sangeeta Malhotra, Bahram
  Mobasher, Keith S.\ Noll, Adam G.\ Riess, Alfred B.\ Schultz, Megan
  L.\ Sosey, Thomas Wheeler, Tommy Wiklind, Chun Xu}

\affil{Space Telescope Science Institute, Baltimore, MD 21218}


%

%
%
%

\paindex{de Jong, R. S.}
\aindex{Arribas, S.}
\aindex{Barker, E.}
\aindex{Bergeron, L. E.}
\aindex{Bohlin, R. C.}
\aindex{Calzetti, D.}
\aindex{Dashevsky, I.}
\aindex{Dickinson, M.}
\aindex{Koekemoer, A. M.}
\aindex{Malhotra, S.}
\aindex{Mobasher, B.}
\aindex{Noll, K. S.}
\aindex{Riess, A. G.}
\aindex{Schultz, A. B.}
\aindex{Sosey, M. L.}
\aindex{Wheeler, T.}
\aindex{Wiklind, T.}
\aindex{Xu, C.}


%
%

\authormark{de Jong et al.}



\begin{abstract}
We provide an overview of the most important calibration aspects of
the NICMOS instrument on board of HST. We describe the performance of
the instrument after the installation of the NICMOS Cooling System,
and show that the behavior of the instrument has become very stable
and predictable. We detail the improvements made to the NICMOS
pipeline and outline plans for future developments. The derivation of
the absolute photometric zero-point calibration is described in
detail. Finally, we describe and quantify a newly discovered
count-rate dependent non-linearity in the NICMOS cameras. This new
non-linearity is distinctly different from the total count dependent
non-linearity that is well known for near-infrared detectors. We show
that the non-linearity has a power law behavior, with pixels with high
count rates detecting slightly more flux than expected for a linear
system, or vice versa, pixels with low count rate detecting slightly
less than expected. The effect has a wavelength dependence with
observations at the shortest wavelengths being the most affected
($\sim$0.05-0.1 mag per dex flux change at $\sim$1 micron, 0.03 mag
per dex at 1.6 micron).

\end{abstract}




\section{Introduction}
NICMOS is currently (December 2005) the second most used science
instrument on board of the Hubble Space Telescope, accounting for
about 25\% of its science observations.  NICMOS has been operating for
more than 3.5 years with the NICMOS Cooling System (NCS) that was
installed in March 2002 during Servicing Mission 3B. With the NCS the
instrument is operating at a very stable temperature, making it easier
to calibrate than in the pre-NCS period, as many instrument
characteristics show a strong temperature dependence. NICMOS has
become over the run of years a more mature instrument on HST with most
of its characteristics well defined and corrected in the standard data
reduction pipeline. However, in recent years NICMOS has been used at
the extremes of its capabilities, revealing new, unexpected
instrumental effects that we are in the process of
calibrating. Examples of extreme use are grism observations of the
$\sim$6$^{th}$ H-mag exo-solar planet host star HD209458 (PI
Gilliland, ID 9642) to the Hubble Ultra Deep Field with galaxies of
about 24$^{th}$ H-mag, a dynamic range of 18 magnitudes!

The organization of the paper is as follows. In section \ref{inst} we
describe the main results of the different calibration programs, in
general monitoring of the instrument under normal circumstances. In
section \ref{pipe} we describe developments of the NICMOS calibration
pipeline and plans for future improvements. Section \ref{phot} gives a
detailed account of the derivation of the photometric zero-point
calibration. Investigations into the recently discovered count-rate
dependent non-linearity are presented in section \ref{nonlin}. Our
further calibration plans are described in section \ref{plans}.

More detailed information on many of the NICMOS calibration efforts
are described in separate contributions to these proceedings.

\section{Instrument}
\label{inst}

Several programs have been executed over the lifetime of NICMOS to
monitor the behavior of key calibrations. One of the most important
programs is the monitoring of the instrument temperature, as many
instrument characteristic depend critical on temperature. We aim to
keep the temperature as close as possible to 77.1 K, adjusting the
cooling rate of the NCS if necessary to account for seasonal
variations and other temperature drifts.

\begin{figure}
\epsfxsize=10cm
\epsfbox[20 160 575 610]{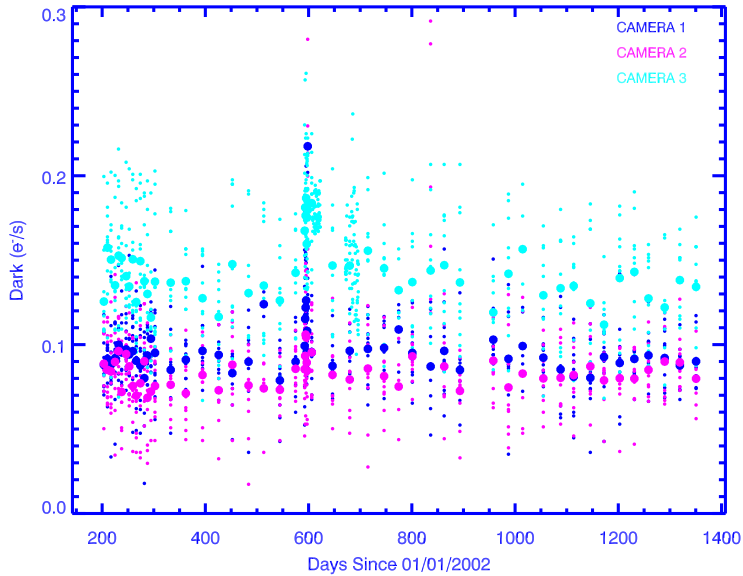}
\caption{Dark count rates in the three different NICMOS cameras.}
\label{dark}
\end{figure}

The dark count rate is monitored at regular intervals and has been
found to be stable (Fig.\,\ref{dark}). The high count rates in
this figure seen near day 600 are due to persistence of Mars
observations. 

\begin{figure}
\epsfysize=12cm
\hspace{0.7cm}\rotatebox{90}{\epsfbox[81 4 583 702]{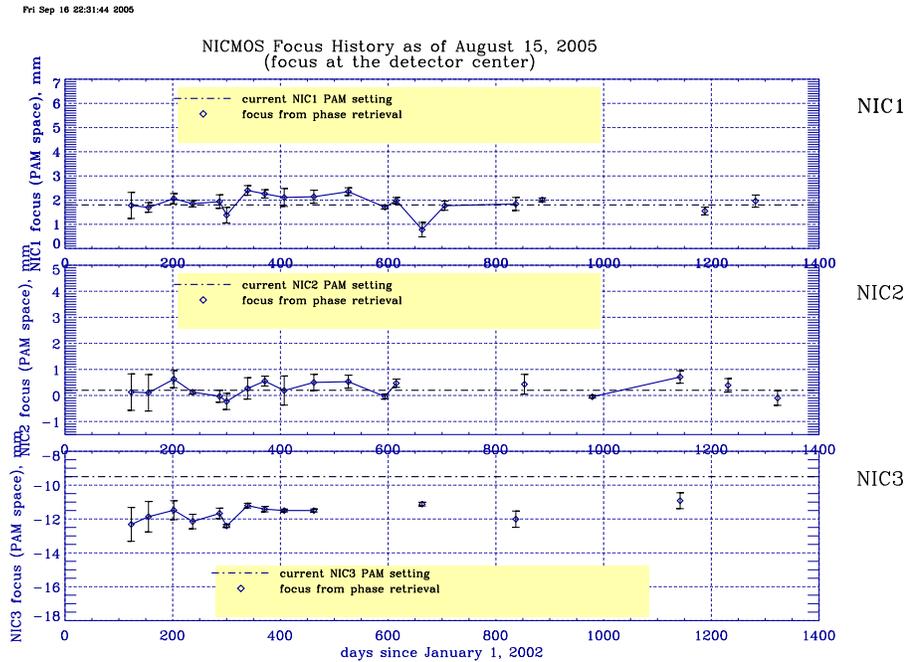}}
\caption{Focus measurements of the three NICMOS cameras. NIC1 and NIC2
  are at their nominal focus setting, NIC3 remains out of reach of the
  PAM focussing mechanism.
  }
\label{focus}
\end{figure}

The focus of the three NICMOS cameras is monitored at regular
intervals using phase retrieval (Fig.\,\ref{focus}). No focus
adjustments have been necessary since the installation of NCS. NIC1
and NIC2 are permanently in focus; NIC3 remains slightly out of focus
due to dewar deformations developed before NCS installation which are
outside the range of corrections possible with the NICMOS focussing
mechanism. No special NIC3 focus campaigns shifting the HST secondary
mirror are planned, as the NIC3 camera is significantly under-sampled
and little is to be gained by such a dedicated campaign.

The count rates detected in the images of the flat field monitoring
program have been very stable, indicating that there has been little
change in sensitivity. However, the general shape of the flat field
has been changing over time since NCS installation. The deviations are
now about 1-3\% from minimum to maximum in NIC1 and NIC3 (NIC2 is less
affected) compared to the pipeline flat fields that were created
after NCS installation. The effect is most severe at the
shorter wavelengths. We are investigating whether these small flat
field variations are the result of temperature changes using the
temperature from bias method (Bergeron, these proceedings). New high
signal-to-noise flat field observations are being planned.

A number of programs were executed to investigate NICMOS performance
in 2-gyro mode (see also Sembach, these proceedings). The NICMOS PSF
shape did not change at the resolution of the NICMOS cameras as
expected. Most critically, the coronographic rejection in the NIC2
camera did not suffer in 2-gyro mode. However, while the NICMOS
coronographic mode is available, it is no longer possible to get two
roll angle coronographic observation in one orbit due to the extra
overhead involved in chancing guide stars under 2-gyro operations.

In June, 2005 new SPARS MULTIACCUM exposure time sequences (SAMP-SEQ)
became operational, replacing the old MIF sequences. The new sequences
are SPARS4, SPARS16, SPARS32 and SPARS128, complementing the already
existing SPARS64 and SPARS256 sequences. These sequences have equal
time steps between each readout. We recommend the use of these SPARS
sequences for most observations, as they provide the most stable
measurements especially in terms of amplifier glow. The
alternative STEP sequences should only be used in situations where one
needs to observe objects with a very large dynamic range in one
observation.

Many of these (and other) calibration investigations have been described in
Instrument Science Reports, which can be found at:
http://www.stsci.edu/hst/nicmos/documents/isrs

\section{Pipeline}
\label{pipe}

Several enhancements are currently underway or have recently been
implemented for NICMOS analysis routines and pipeline-related
software. The MultiDrizzle software (Koekemoer et al. 2002, 2005)
available within Pyraf has been extended to enable fully automated
combination of calibrated NICMOS images, which can be provided either
as a list of exposures or as a NICMOS association table.  We have also
distributed the first Pyraf release of the "SAAclean" task (Barker et
al. 2005) which is based on the IDL algorithm (Bergeron and Dickinson
2003) to remove residual flux from pixels impacted by cosmic rays
accumulated during passage through the South Atlantic Anomaly. Testing
is currently underway on both MultiDrizzle and SAAclean to prepare
them for eventual inclusion into the automatic processing carried out
by the HST archive pipeline.

Development is also currently proceeding on a task to remove the
cross-talk effect known as "Mr. Staypuft", where flux from bright
sources on one quadrant can be seen to propagate to pixels in
corresponding locations on the other quadrants. This task will be
released to the community in an upcoming Pyraf release, and may
eventually also be incorporated as part of the HST pipeline.  Future
work will include software to determine the NICMOS detector
temperature from bias and voltage measurements, as well as improved
amplifier glow and pedestal correction software which will most likely
make use of the improved temperature measurements.

\section{Photometry zero-point}
\label{phot}

Initial absolute photometric calibrations for NICMOS were obtained
during SMOV in July 1997. These preliminary results used only a few
filters in each camera to establish initial corrections from the
predicted ground-based vacuum measurements. These measurements obtained
accuracies between 10 and 15 percent.  Later observations improved the
calibration to about the 5\% level in all cameras and filters by May
1998. However, revised ground-based photometry from Persson et
al.\ 1998 suggested that calibrations based on the these
CDBS (Calibration Data Base System) NICMOS photometric standard stars
observations might be off by as much as 5 - 14 percent in some
filters, due to an 0.1 magnitude discrepancy with the ground-based
photometry for the solar analog and white dwarf used for calibration.

With the installation of the NCS the NICMOS detectors are operating at
dramatically higher temperature with an associated strong change in
sensitivity.  Furthermore, the limited accuracy of previous analyses,
new reduction routines, improvements to the pipeline calibrations and
calibration images, and a better understanding of the instrument born
anomalies encouraged a complete re-analysis of all available NICMOS
photometric calibration data. The following paragraphs will discuss
calculation of new aperture corrections for each camera/filter
combination, improvements of the spectrophotometric standard star
spectra used for calibration and an assessment of the overall
photometric stability during the full lifetime of the instrument.

All standard star observations were reduced with the latest version of
the calibration pipeline and reference files. In post-pipeline
processing quadrant dependent bias was removed with {\it pedsky} and
residual readout shading was removed by 1D fitting of the sky. 

\subsection{Aperture Corrections}

\begin{figure}
\epsfxsize=12cm
\epsfbox[42 180 555 550]{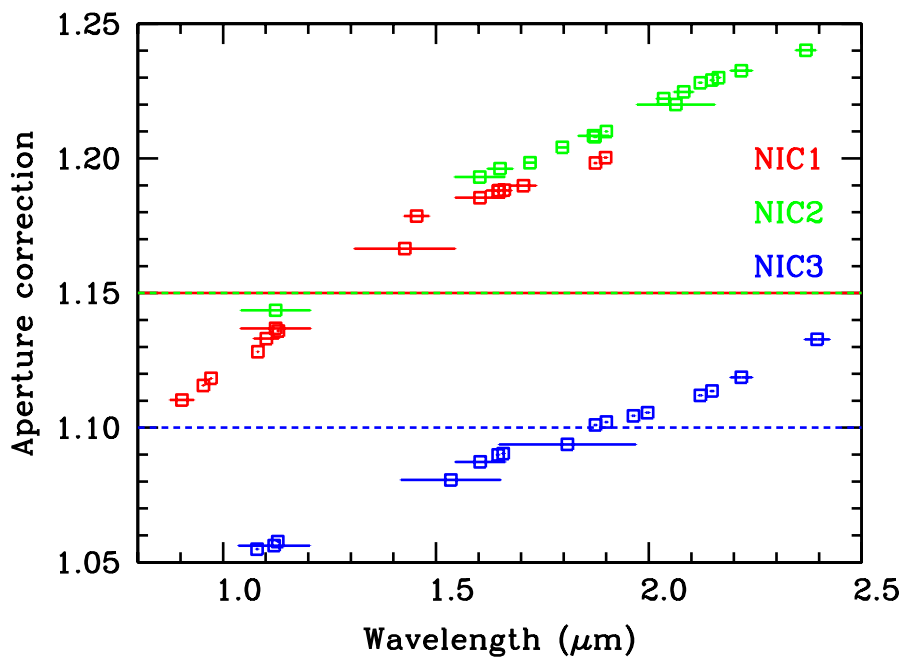}
\caption{Comparison of the old aperture corrections (dashed lines)
  with the new aperture corrections (squares, the filter widths are
  indicated by the horizontal line).
\label{apcorr}
}
\end{figure}

Previous calibrations used a single aperture correction regardless of
the filter element to correct the fixed aperture standard star
photometry to an infinite aperture. While this can be used
consistently for point source observations, the wavelength dependence
of the PSF makes this incorrect for calibrating extended sources and
makes comparisons with ground-based observations difficult. New,
wavelength dependent, aperture corrections for each of the cameras
have been calculated using the TinyTim PSF modeling software.
Figure\,\ref{apcorr} shows the new aperture corrections compared to
the previous fixed corrections. The new aperture corrections, along
with the aperture sizes and sky annuli used, are tabulated on the
NICMOS photometry web site.

\subsection{Absolute Spectrophotometric Standards}

The photometric calibration keywords are derived from the comparison
of the measured NICMOS count rates for the standard star observations
to the spectral flux density of the standard star averaged over the
NICMOS bandpass. There are no ground-based spectrophotometric
observations of standard stars with complete coverage over the NICMOS
wavelength range, and therefore we must use accurately calibrated
"surrogate spectra" instead in our comparison.  This was the
motivation for using solar analog and white dwarf standards for the
NICMOS photometric calibration.  The absolute spectral energy
distribution of the Sun is well known (see Colina, Bohlin \& Castelli
1996 and references therein), and thus can be scaled reliably to
represent the spectrophotometry for solar analog standard stars like
P330E.  DA white dwarfs like G191B2B have relatively simple stellar
atmospheres, and considerable effort has gone into accurately modeling
these and comparing them to UV-through-optical spectrophotometry
(Bohlin, Colina \& Finley 1995; Bohlin 1996; Bohlin 2000).

As described by Colina \& Bohlin 1997, the infrared spectrum of P330E
is represented by the solar spectrum from Colina, Bohlin \& Castelli
1996.  The white dwarf G191B2B is represented by an LTE model
calculated by D. Finley (described in Bohlin 2000).  These
spectrophotometric models are then normalized using ground-based
photometry of the NICMOS standard stars.  Persson et al.\ (1998) have
obtained ground-based JHK photometry for a large set of faint infrared
standard stars, including the HST/NICMOS solar analog standards.
Persson (private communication) also observed G191B2B as part of the
same program.

In order to normalize the standard star spectral models, we must
convert Persson's JHK magnitudes to absolute flux density units.
Campins, Rieke \& Lebofsky (1985) provide an absolute infrared flux
calibration scale using a solar analog method.  However, the effective
wavelengths and bandwidths of their JHK filters (which we will refer
to as the Arizona system, where Vega is defined to have $J$=$H$=$K$=0.02)
differ somewhat from those used by Persson et al.\ (calibrated to the
CIT system, where Vega is defined to have $J$=$H$=$K$=0.0).  In order to
shift the Campins et al.\ absolute calibration to the Persson et
al.\ bandpasses, we have used an ATLAS 9 atmosphere model for Vega.
This model is not used for any absolute calibration, but simply to
compute flux density ratios for Vega between the Arizona and Persson
et al.\ bandpasses.  These are then used to convert the Campins et
al.\ Vega flux densities to the Persson et al.\ bandpasses, and hence to
provide the absolute flux density calibration for the Persson et
al.\ measurements.  In this way, $m$=0 is calibrated to be 1626, 1056,
and 658 Jy for the Persson et al.\ {\em JHK} bandpasses, respectively.

The P330E and G191B2B spectrophotometric models are then synthetically
integrated through the Persson et al.\ {\em JHK} passbands, and the
bandpass-averaged flux densities are converted to magnitudes for
comparison to the ground-based photometry.  This comparison indicates
that the Colina \& Bohlin (1997) P330E model requires an average flux
re-normalization of +7\% to match the {\em JHK} photometry at the 0.01 mag
level.  For G191B2B, the synthetic and ground-based
photometry agree precisely (0.002 mag) at $J$ and within 0.02 mags at $K$,
but differ by 0.053 mags at $H$.  We adopt the Bohlin (2000) G191B2B model
without change, but note this possible discrepancy near 1.6 microns.

\subsection{Zero-point Calibration}

\begin{figure}
\epsfxsize=12cm
\epsfbox[52 190 545 550]{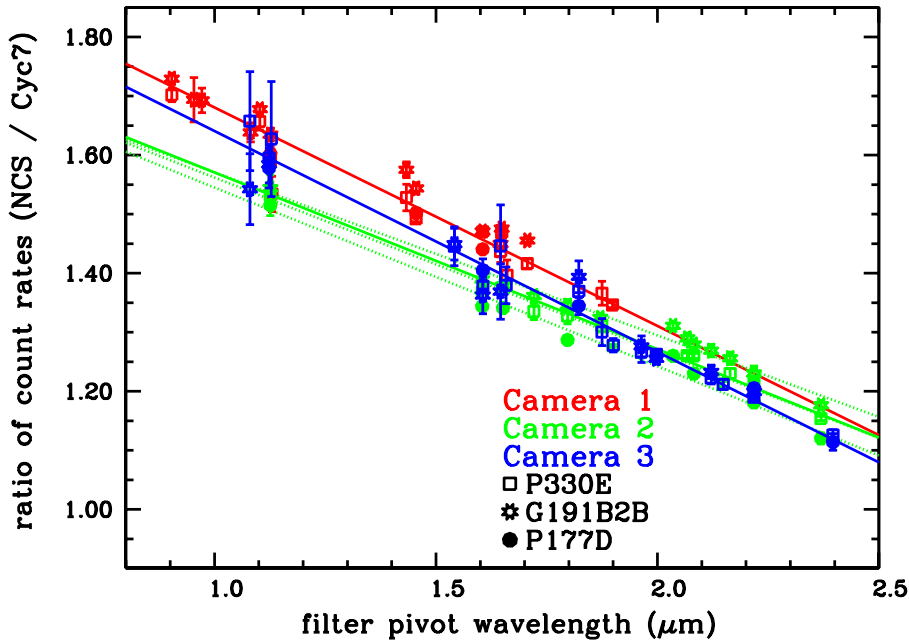}
\caption{The relative change in count rate of standard stars from
  Cycle 7 to post-NCS observations. The lines are linear fits to the
  data of the three cameras
\label{cr_comp}
}
\end{figure}

For each standard star we determined the average aperture count rate
for all NIC1, NIC2, and NIC3 filters for each dither position and the
many repeat observations. Observations near bad pixels or other
outliers were removed from the average. The observed, aperture
corrected, average total count rates for each star in each filter were
compared with the predicted total count rates obtained using SYNPHOT
synthetic photometry package\footnote{available at
  http://www.stsci.edu/resources/software\_hardware/stsdas/synphot} on
the spectra of G191B2B and P330E. We used the {\em
 calcphot} program to calculate the effective stimulus of a source
with a flat spectrum in $f_\nu$. We used the ground-based filter
shapes and detector quantum efficiency QE curve for the Cycle 7
data. The QE curve was modified for the post-NCS change in sensitivity
by multiplying the QE with a linear correction that was determined
from comparing pre-NCS to post-NCS count rates of standard stars
(Fig.\,\ref{cr_comp}). Comparing the thus calculated effective
stimulus to the observed average count rates results in the PHOTFNU
keywords. Similar calculations provide the PHOTFLAM, PHOTPLAM and
PHOTBW values found in NICMOS image headers, as described in the
SYNPHOT manual and in Sirianni et al.~(2005). The new calibration
values have been automatically provided in the image headers retrieved
from the HST archive since June 2004. The new filter throughput files
were made available in the CDBS in December 2005. The latest
calibration values can also be found at the NICMOS photometry web
pages: http://www.stsci.edu/hst/nicmos/performance/photometry .

\subsection{Photometric Stability}

\begin{figure}
\mbox{
\epsfxsize=7.5cm
\epsfbox{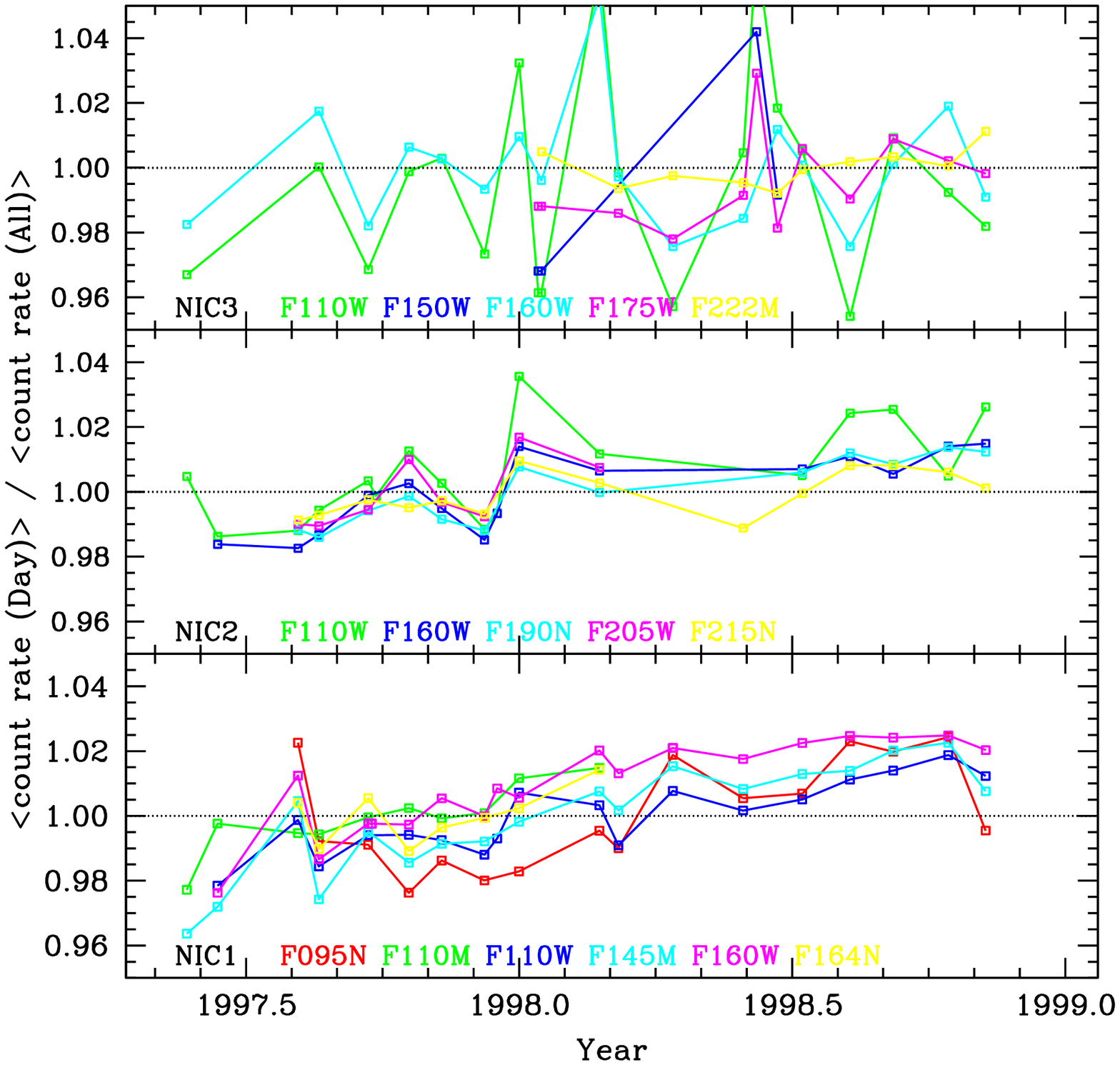}
\epsfxsize=7.5cm
\epsfbox{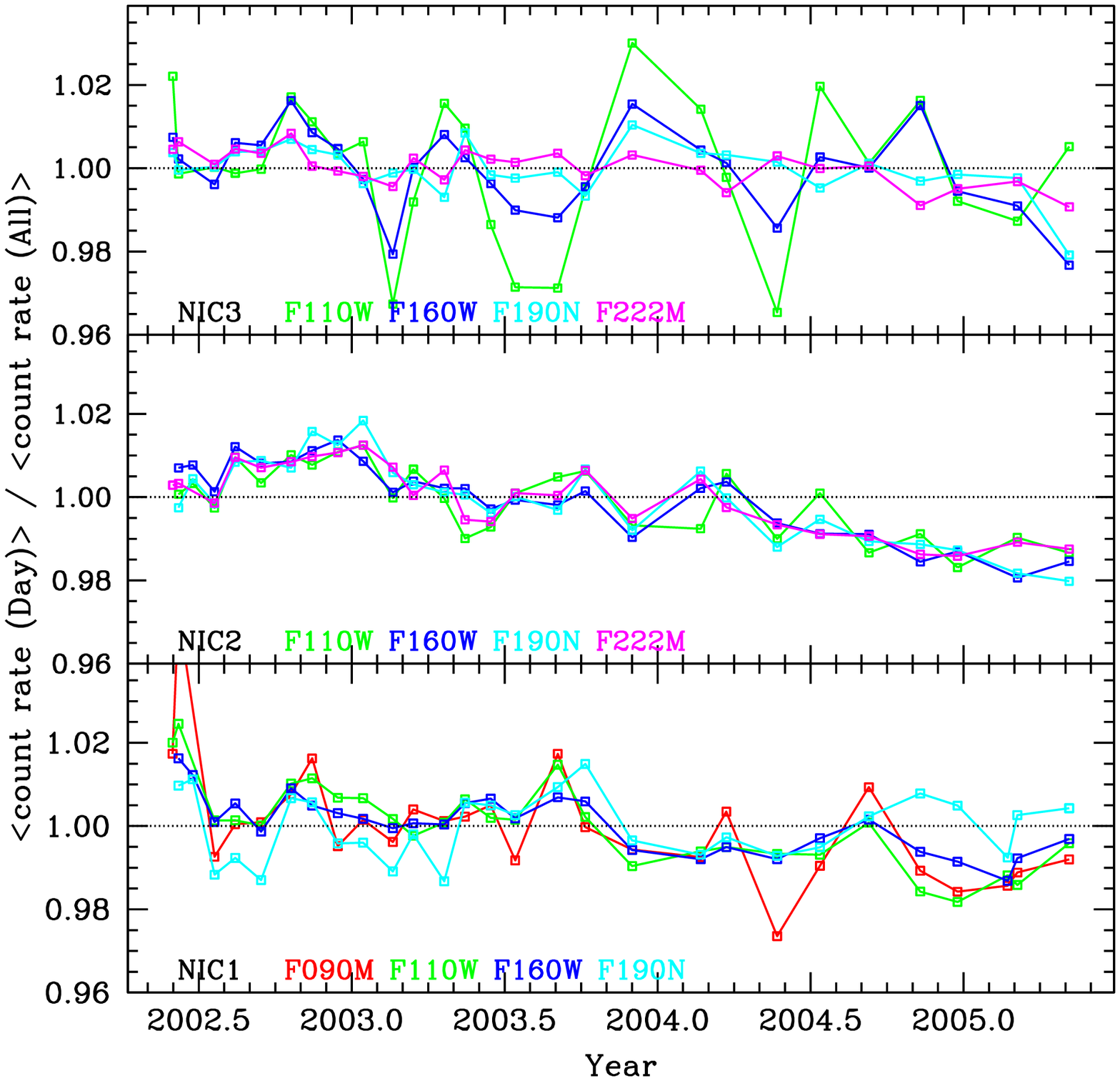}
}
\caption{Evolution of the P330E standard star count rates with left
  the Cycle 7 data and right the post-NCS data.
\label{photmon}
}
\end{figure}

Photometric stability of NICMOS was investigated using all
observations of Solar analog star P330E used in the photometric
monitoring program (Fig.\,\ref{photmon}). The Cycle 7 data show the
clear increase in sensitivity due to the increase in temperature while
the solid nitrogen evaporated. After temperature correction there is a
slight decrease in sensitivity. The post-NCS data show clear decrease
in sensitivity in NIC2, a downward trend in NIC1, but the NIC3 data is
too noisy due to intra-pixel sensitivity variations to tell whether
there is any downward trend. The cause of this decrease in standard
star sensitivity is not clear yet, as it is not matched in the flat
field lamp monitoring data. However, similar downward trends are seen
in the few repeat observations of G191B2B.

\section{Non-linearity}
\label{nonlin}

\begin{figure}[th]
\epsfysize=13cm
\rotatebox{90}{\epsfbox{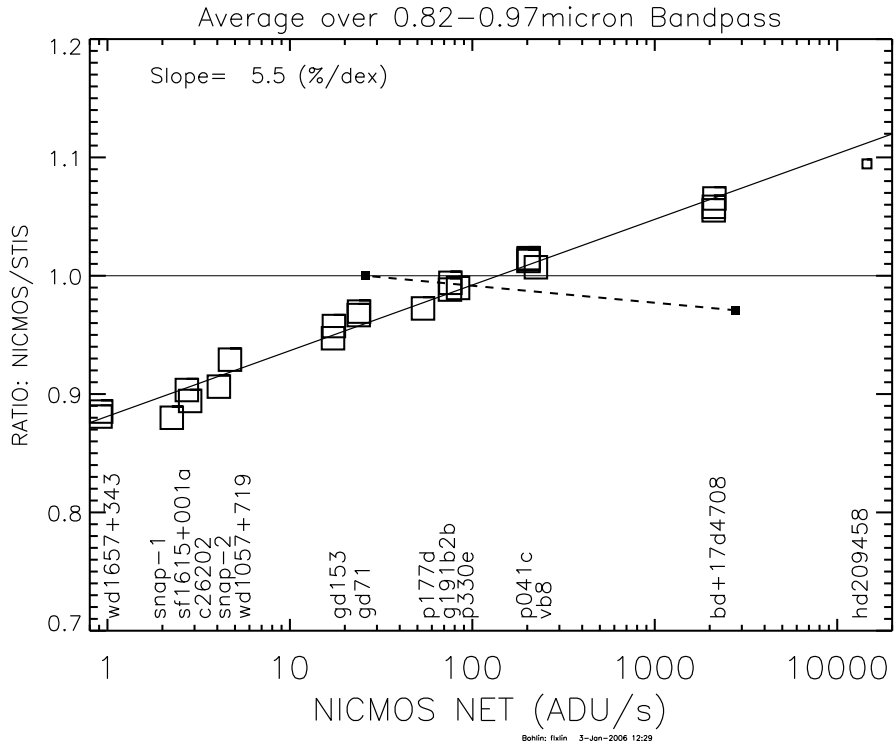}}
\caption{Ratio of NICMOS grism fluxes to the STIS measurements
  averaged over the 0.82--0.97$\mu$m overlap range, normalized to the
  P330E ratio (big open squares) and least square fit (solid
  line). There is a non-linearity of about 5.5\% per dex or a total of
  23\% over the 4 dex dynamic range in observed response.  The bright
  star HD209458 was observed with NICMOS in a defocused mode and not
  included in the fit. Small filled squares connected by a dashed
  line---ACS grism fluxes compared to STIS in the same
  0.82--0.97$\mu$m band. Over a dynamic range of $\sim$100$\times$
  between GD153 and BD+174708, the CCD detectors on STIS and the HRC
  in ACS measure the same relative flux to within 2\%.
\label{nonlinfig}
}
\end{figure}

In a recent analysis of NICMOS, STIS and ACS spectral data Bohlin et
al.~(2005) found that NICMOS shows a systematic count rate dependent
non-linearity, primarily at the shorter wavelengths
(Fig.\,\ref{nonlinfig}). The same spectra show a similar non-linearity
when compared to ACS photometry. This count-rate dependent
non-linearity is distinctly different from the normal non-linearity of
near-infrared detectors that depends on the total counts, not on the
count rate. The total count non-linearity is well understood and
corrected in the NICMOS pipeline.

The non-linearity is such that at high count rate there are more
counts than expected, at low count rates less than expected, compared
to intermediate count rates. The non-linearity shows no sign of
turnover over the full 4 orders of magnitude measured
(Fig.\,\ref{nonlinfig}), and hence we cannot say with confidence that we
are missing photon detections at the faint end or are getting extra
detections at the bright end. The non-linearity is well modeled by a
power law.

The count rates of the NICMOS spectra show in general good agreement
with the NICMOS photometry count rates of the same objects, indicating
that the NICMOS system is internally consistent and that it is not the
spectral data reduction that is at fault. A few more indications have
been found that NICMOS suffers from a non-linearity dependent on the
incoming flux: 1) narrowband filters at the shorter wavelengths
required larger in flight corrections from their ground-based determined
throughputs than the broadband filters, 2) high redshift supernova
fluxes are slightly fainter in F110W than expected based on their ACS
fluxes and well tested SN models (Adam Riess, private communication),
and 3) galaxies in the HUDF are slightly fainter than expected based
on ACS and ground-based J\&K magnitudes combined with SED modeling
(Mobasher \& Riess 2005; Coe et al., these proceedings).

However, all these lines of evidence rely on modeling of filter
throughputs and/or spectral modeling of sources. Here we describe a
test that depends on the change in incoming flux on the detector
alone. NICMOS is a shutterless instrument and observes the sky while
obtaining calibration flat fields using its internal lamps. The same
object can be observed with an artificially increased flux and the
count rate for objects can be compared with and without extra
lamplight. For a fully linear system adding a background flux should
not enhance the flux in the object, but any flux dependent
non-linearity is revealed immediately when subtracting lamp-off images
from lamp-on images.

\begin{figure}
\epsfxsize=\textwidth
\epsfbox{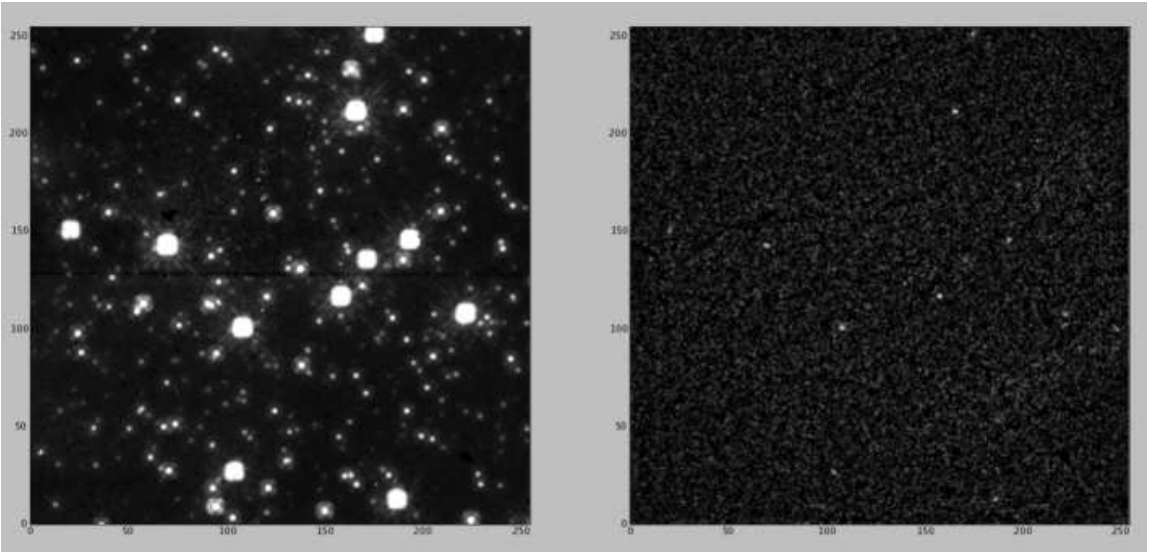}
\caption{(Left) NIC1 F110W lamp-off image of NGC\,1850. (Right) NIC1
  F110W lamp-on minus lamp-off image. Bright stars are clearly not
  well subtracted and leave residual flux as expected for count rate
  dependent non-linearity.
\label{lampres}
}
\end{figure}

\begin{figure}
\epsfxsize=16cm
\epsfbox{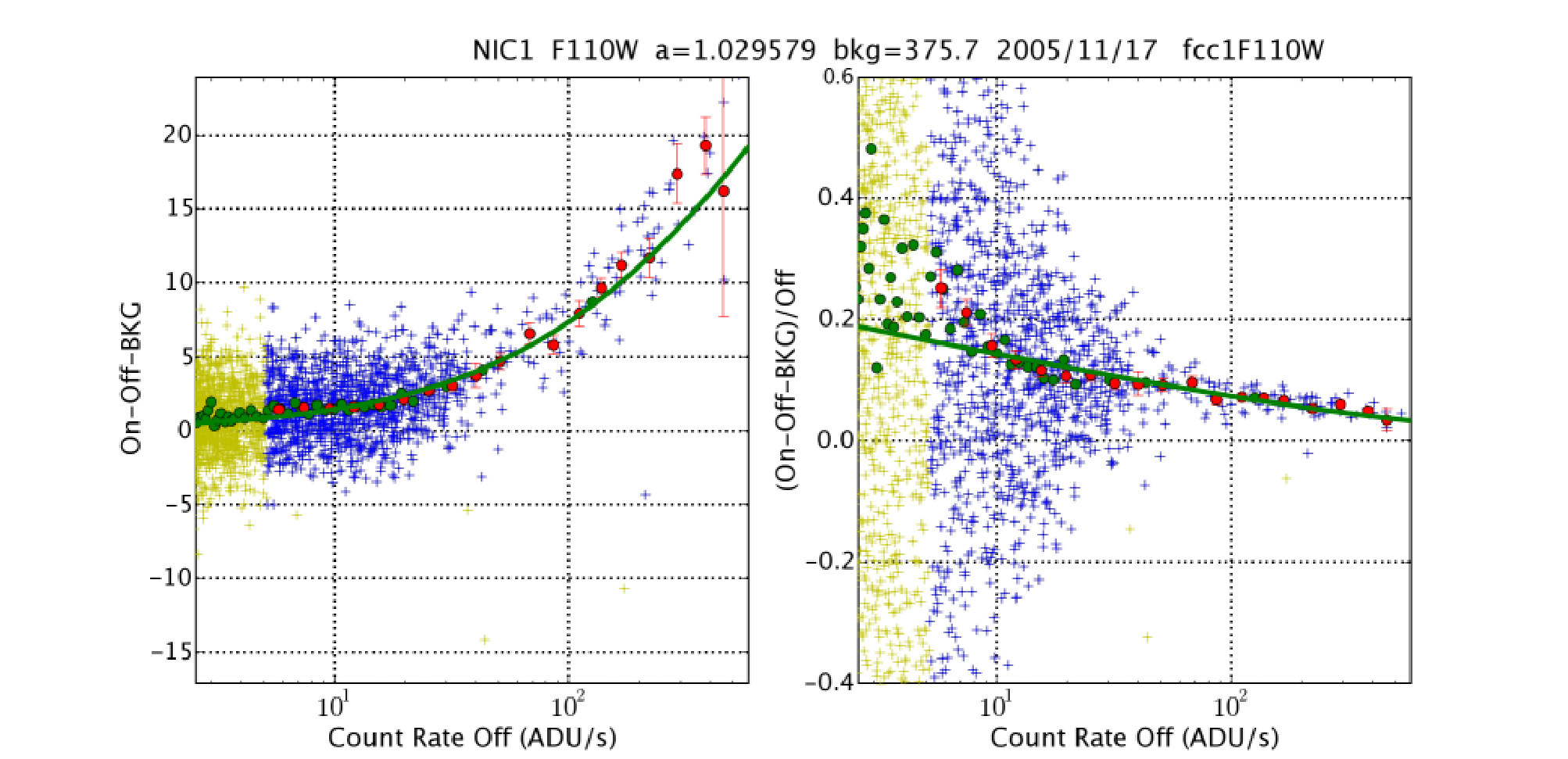}
\epsfxsize=16cm
\epsfbox{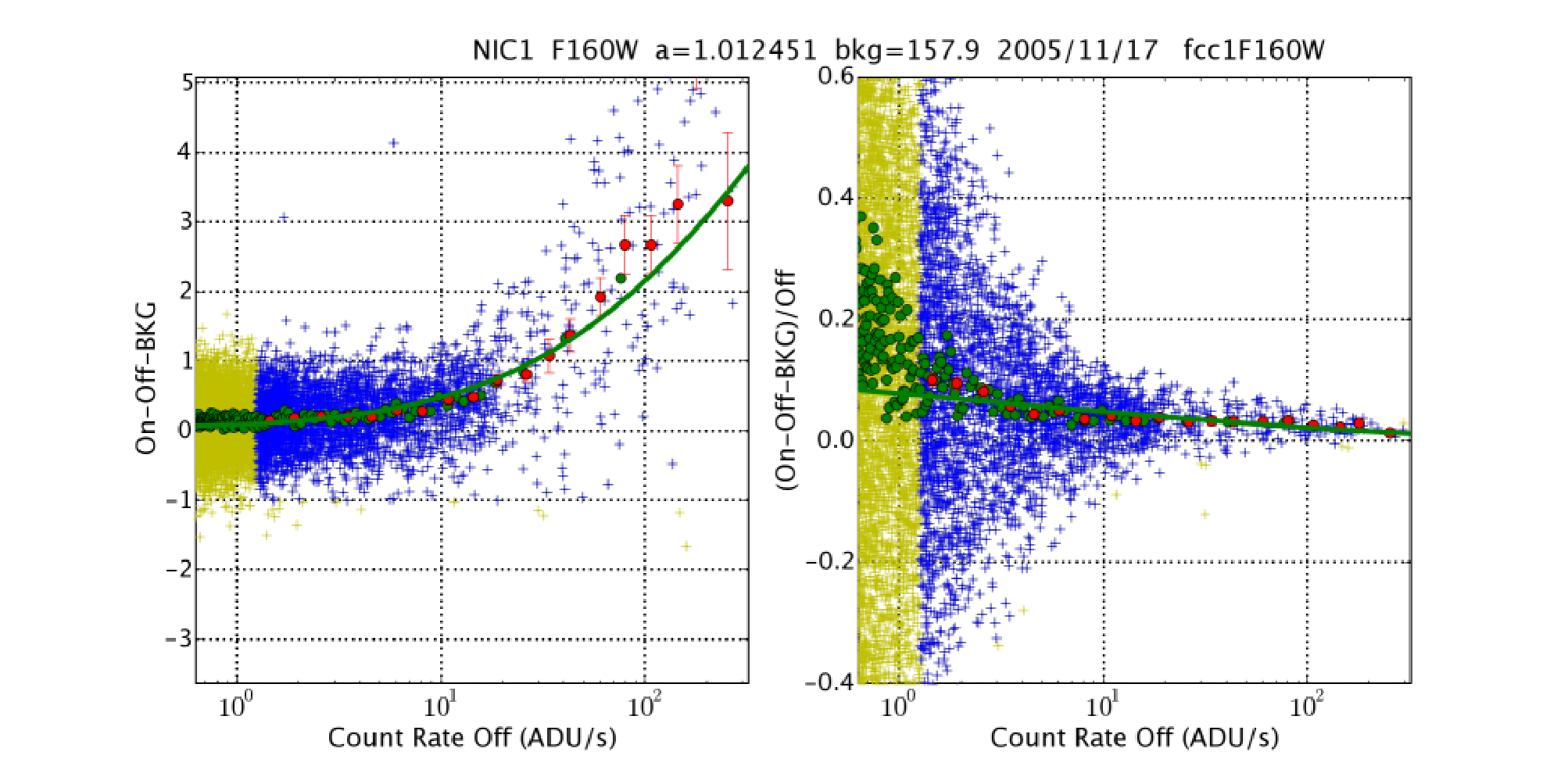}
\caption{ The absolute and relative difference in lamp-on minus
  lamp-off count rates as function of the lamp-off count rates on a
  pixel-by-pixel basis. The yellow + symbols are for all data, the
  blue + symbols for the data used in the fit (bad and low S/N points
  filtered). The green circles are the averages in bins of 50 pixels
  in ascending lamp-off count rate for all pixels. The red circles are
  binned averages in 30 equal logarithmic steps in lamp-off count rate
  for the selected pixels. The green lines are the fitted
  non-linearity functions, with the fitted $\alpha$ parameter labeled
  at the top. Left is the absolute difference
  ($cr_{on}-cr_{off}-cr_{lamp}$), right is the relative difference
  ($cr_{on}-cr_{off}-cr_{lamp}$)/ $cr_{off}$. While the brightest
  points have the largest flux change in absolute sense and are easy
  to measure above the noise (and are not due to background
  subtraction errors), the fainter points change relatively the most
  and have larger calibration errors relative to bright standard
  stars. Top) NIC1 camera, F110W filter. Bottom) NIC1, F160W filter.
\label{lamponoff}
}
\end{figure}

Star cluster NGC\,1850 was observed in the Cycle 14 calibration
program in a sequence of lamp-off, lamp-on, and lamp-off using the
same telescope pointing and without changing the exposure
sequence. Similar observations taken in Cycle 7 for a different
purpose were also analyzed. The data were analyzed under the
assumption that a power law can model the non-linearity:
\[
cr(x,y) \propto (f_{tot}(x,y))^\alpha, 
\]
with $cr(x,y)$ the measured count rate in ADU/s and $f_{tot}(x,y)$ the
total flux falling on a detector pixel at $(x,y)$. For a non-linearity
of $\sim$5\% per dex this corresponds to $\alpha$$\sim$1.02. In magnitudes
we have an offset of $\Delta m=2.5(\alpha-1)$ per dex change in
incident flux. When we subtract the lamp-off from the lamp-on
observation we expect to see positive residuals at positions where
there are objects if $\alpha$$>$1:
\[
cr_{on}-cr_{off}
\propto(f_{obj}+f_{sky}+f_{lamp})^\alpha-(f_{obj}+f_{sky})^\alpha 
\sim (f_{obj}+f_{lamp})^\alpha-(f_{obj})^\alpha,
\]
where it is assumed that the sky flux is small compared to the other
fluxes. Such image residuals are shown in Figure\,\ref{lampres}. The
absolute boost in measured count rate is largest for bright objects,
but the relative increase in measured count rate is larger for lower
object fluxes, because the relative increase in flux by switching on
the lamp is much larger. However, at low count rates the noise
dramatically increases and we have to average many points to see the
effect. This is shown in Figure\,\ref{lamponoff}, where we plot both
the absolute and relative count rate increase due to the lamp
background.

\begin{table}
\begin{center}
\begin{tabular}{ccccc}
\hline
  Date	  & Camera&	Filter&	$\alpha$& $\Delta$m/dex\\
\hline
\multicolumn{5}{c}{Cycle 7}\\
1998/02/18&	2&	F110W&	1.022 $\pm$ 0.001&	0.055 $\pm$ 0.003\\
1998/04/17&	2&	F110W&	1.025 $\pm$ 0.004&	0.063 $\pm$ 0.010\\
1998/06/04&	2&	F110W&	1.023 $\pm$ 0.001&	0.059 $\pm$ 0.002\\
1998/08/06&	2&	F110W&	1.024 $\pm$ 0.001&	0.061 $\pm$ 0.002\\
1998/09/24&	2&	F110W&	1.022 $\pm$ 0.001&	0.054 $\pm$ 0.002\\
\hline
\multicolumn{5}{c}{Cycle 14}\\
2005/11/17&	1&	F090M&	1.040 $\pm$ 0.003&	0.101 $\pm$ 0.008\\
2005/11/17&	1&	F110W&	1.030 $\pm$ 0.003&	0.074 $\pm$ 0.009\\
2005/11/17&	1&	F160W&	1.012 $\pm$ 0.002&	0.031 $\pm$ 0.006\\
2005/11/17&	2&	F110W&	1.025 $\pm$ 0.002&	0.063 $\pm$ 0.006\\
2005/11/17&	2&	F160W&	1.012 $\pm$ 0.006&	0.029 $\pm$ 0.015\\
2005/11/17&	2&	F187W&	1.005 $\pm$ 0.004&	0.013 $\pm$ 0.009\\
\hline
\end{tabular}
\end{center}

\caption{Measured $\alpha$ and $\Delta m$ values in Cycle 7 and 14
\label{nonlintab}
}
\end{table}

The fitted non-linearity functions are overplotted in
Figure\,\ref{lamponoff} and the measured $\alpha$ values are tabulated
in Table\,\ref{nonlintab}. A number of points can immediately be taken
from the Table. NICMOS has a significant count-rate dependent
non-linearity, becoming more severe at shorter wavelengths. This is a
different non-linearity from the well-known total count dependent
non-linearity. The non-linearity in NIC1 and NIC2 amounts to 0.06-0.10
mag offset per dex change in incident flux for the shortest wavelength
(F090M and F110W), about 0.03 mag/dex at F160W and less than that at
longer wavelengths. These corrections are larger than predicted from
the Bohlin et al.~(2005) NIC3 grism results, which may point to
intrinsic detector differences or might be the result of a different
analysis method. The non-linearity seems to have changed very little
from Cycle 7 to Cycle 14 (in F110W NIC2), and hence is unlikely to
depend on detector temperature. The fact that there is a wavelength
dependence to the effect in the lamp off/on/off test and that this
trend quantitatively agrees with the grism observations strongly
argues against this being the result of a data reduction error and
that the cause is intrinsic to the measurement.

To what extend NICMOS photometry is affected by the non-linearity
depends on the wavelength of the observations (i.e. the $\alpha$
parameter), whether the object is a point source or extended, and on
the count rate of the sky background (as the count rate will never go
below the sky level and hence the non-linearity will level off, even if
the sources have lower count rates).  Given that the NICMOS standard
stars are of about the 12th magnitude, the maximum expected offset for
the Hubble UDF for example is about 0.15-0.2 mag at 22 F110W AB-mag, where
the objects are comparable to or below the sky count level (see
contribution by Mobasher, Thompson and Coe in these proceedings for
further analysis of the HUDF).

The NICMOS team has as of yet not found a physical explanation for the
count rate non-linearity. However, with these measurements we can
start to calculate corrections to point source photometry. Software is
being developed for more complicated cases with extended sources or
objects close to the sky background level. Eventually corrections will
be incorporated in the NICMOS calibration pipeline. Non-linearity lamp
on/off calibration observations for all regularly used filter/camera
combinations are currently being planned. More information on the NICMOS non-linearity and how to correct for it can be found at:
http://www.stsci.edu/hst/nicmos/performance/anomalies/nonlinearity.html .

\section{Calibration plans}
\label{plans}

The NICMOS team is currently implementing a calibration plan that
addresses several of the anomalies observed above. New, high
signal-to-noise flat field observations will be obtained for all
filters regularly used. Dither observations with NIC1 have been
obtained of star cluster NGC\,1850 to investigate the low frequency
flat field variations seen in the photometric monitoring data. Lamp
on/off observations of star fields will be obtained in all regularly
used camera/filter combinations to investigate the count-rate
dependent non-linearity. Details of the NICMOS calibration plan for
Cycle 14 can be found in the contribution by Arribas to these
proceedings.

\end{document}